# The oral tolerance as a complex network phenomenon

*Miranda, Pedro Jeferson [1]; Delgobo, Murilo [3]; Favero, Giovanni Marino [3]; Paludo, Kátia Sabrina [3]; Baptista, Murilo [2]; de Souza Pinto, Sandro Ely [1,2]*

1. Departamento de Física, Universidade Estadual de Ponta Grossa, 84030-900, Brazil
2. Institute for Complex Systems and Mathematical Biology, SUPA, University of Aberdeen, Aberdeen, United Kingdom
3. Departamento de Biologia, Universidade Estadual de Ponta Grossa, 84030-900, Brazil

**Abstract -** The phenomenon of oral tolerance refers to a local and systemic state of tolerance, induced in the gut associated lymphoid tissues (GALT), after its exposure to innocuous antigens, such as food proteins. While recent findings shed light in the cellular and molecular basis of oral tolerance, the network of interactions between the components mediating oral tolerance has not been investigated yet. Our work brings a complex systems theory approach, aiming to identify the contribution of each element in an oral tolerance network. We also propose a model that allows dynamical plus topological quantifying which must encompass functional responses as the local host involved on the oral tolerance. To keep track of reality of our model, we test "knock-out" (KO) of immunological components (i. e. silencing a vertex) and see how it diverges when the system is topologically "healthy". The results from these simulated KOs are then compared to real molecular knock-outs. To infer from these processing we apply a new implementation of a random walk algorithm for directed graphs, which ultimately generate statistical quantities provided by the dynamical behavior of the simulated KOs. It was observed that the knockout of vertex $CD103^+$ caused the greatest impact on network *standard flux*, resulting in a *mean relative deviation* of 0,65. The knockout of vertexes $iTregFoxP3^+$ and Tr1 lead to a *mean relative deviation* of 0,37 and 0,38 respectively, and for TGF-β and RA, 0,29 and 0,23 respectively. In a brief analysis, the results obtained correspond to biological data, where DCs $CD103^+$ plays critical roles in the generation of Tregs, through mechanisms dependent on TGF-β and RA. Our model addresses both topological proprieties and dynamical relations. The construction of a qualitative dynamic model for oral tolerance could reflect empirical observations, through the standard flux results and relative error based on individual knockout.

## Introduction

The adaptive immune response (AIR) constitutes a remarkable characteristic of jawed vertebrate organisms, given its plasticity and specificity of systemic responses. In general, it is mostly composed of antibody secreting cells, also known as B cells, and by a set of specialized T cells, which modulates distinct functions on immune response [1]. As plasticity and specificity of AIR we mean that while it has the ability to eliminate harmful pathogens, it must also generate a sort of tolerance to benign epitopes to avoid immunological responses against self-components of the system. These components come into a variety; mainly as self-structures of the system, but also as out-particles, important to the system, for example, food particles [2].

The thymus is the responsible organ for T cell development and maturation, as the self and non-self peptides are processed by antigen presenting cells (APCs) which is bound to major histocompatibility complex (MHC) proteins; the T cell receptors (TCRs) displays a highly variable region for antigen recognition, generate by stochastic rearrangement of relevant genes. T cells which recognize self-MHC molecules receive survival signals and are expanded, in a process called positive selection. However, T cells which recognize self-peptides may undergo negative selection (cell death) or become naturally occurring regulatory T cells (nTreg) [3, 4]. Naïve T cells (T cells that had not encountered its cognate antigen) may leave the thymus and mediate peripheral tolerance to environmental antigens in mucosal sites, such as the oral cavity and the gut. Following exposure to the antigen, naïve T cells become induced regulatory T cells (iTreg). In contrast to nTregs, iTregs displays transient phenotype, depending on the milieu conditions they are present [5].

The phenomenon of oral tolerance refers to a local and systemic state of tolerance, induced in the gut associated lymphoid tissues (GALT), after its exposure to innocuous antigens, such as food proteins [6]. APCs resident in intestinal *lamina propria* (LP) capture antigens from the lumen and migrate to mesenteric lymph nodes (mLN), where they drive T cell differentiation. Tregs generated in the mLN may return to the LP or enter bloodstream via spleen, where they promote the systemic effects of oral tolerance [7]. While Immunology has made solid advances in terms of defining the genetics, molecular and cellular components involved in oral tolerance phenomena [8, 9], the network of interactions between these components has not been investigated yet.

Actually, little information is available how all these molecular processing occurs simultaneously.

As matter of fact, the oral tolerance relies on the complex interactions of immune components in a unique microenvironment (GALT) [10]. It is of importance to shed light onto this problem under the paradigm of the complex network theory. This proportionates a new way to assess properly the interrelationship of parts, which shall, heuristically, generate answers that are not accessible to experimental analysis alone [17]. To make this assessment viable, we propose a model that allows dynamical plus topological quantifying which must encompass functional responses as the local host involved on the oral tolerance. To keep track of reality of our model, we test "knock-out" (KO) of immunological components (i. e. silencing a vertex) and see how it diverges when the system is topologically "healthy". The results from these simulated KOs are then compared to real molecular knock-outs. To infer from these processing we apply a new implementation of a random walk algorithm for directed graphs, which ultimately generate statistical quantities provided by the dynamical behavior of the simulated KOs.

**The oral tolerance as a complex network phenomenon**

To approach the oral tolerance phenomenon via complex network, it is necessary to change the paradigm on how this phenomenon must occurs ontologically [18]. The first pragmatic consideration taken from this paradigm shift is the association of our phenomenon as a systemic response to a stimulus. This very initial trigger must unchain a process which starts a set of special interactions of immunological components. It, reductively, must ensure a minimal quantity of differentiated T cell (Tregs). The phenomenon *per se* should rely on how these components are related to each other and when these interactions occur. Given these premises, we propose a network which captures interactions and a time dependant quantity based on a "simulated stimulus". To make this possible we make use of the random walk in complex networks which, ultimately, shall define our quantifiers.

The interaction network of oral tolerance were built from the available literature and adapted to our system goal. Immune components such as lymphocytes, cytokines and antigens were represented as network's vertices. The interactions, regulatory relationships and transformations among components were described as directed edges, starting from the source vertex and ending on the target vertex. Some foreign peptides (antigens) can resist both the low pH of the gastric fluid and proteolytic enzyme hydrolysis, reaching the small intestine lumen as large immunogenic peptides or intact proteins [11]. These antigens can be complexed to IgG and IgA in the lumen, and transported to intestinal *lamina propria* (LP) through neonatal Fc receptor (FcRn) and transferrin receptor (CD71) respectively [11, 12]. Enterocytes, that are intestine absorptive cells, plays a critical role in antigens capture. Small molecules (<600 Da) may pass through enterocytes' tight junctions. Enterocytes can fuse partially degraded proteins to MHC II compartments and deliver it to LP in form of exosomes [7, 13]. Antigens can also be transported to Peyer's Patches through M cells. [14] A particular population of dendritic cells (DCs) residing in the LP (DCs $CD103^+$) receives and loads the antigens from intestinal lumen. These cells migrate to mesenteric lymph nodes (mLN) where they induce the differentiation of *naïve* CD4 T cells to T regulatory cells. Basically, DCs $CD103^+$ present antigens via MHC II combined with transforming growth factor β (TGF-β) and retinoic acid (RA), favoring the differentiation of iTregsFoxP3$^+$ [8, 15]. RA imprints gut-homing molecules on Tregs, so they can return to LP and proliferate in an IL-10 dependent mechanism. iTregFoxP3$^+$ cells induces the secretion of IL-27 by DCs $CD11b^+$, which enhances the production of IL-10 by type I regulatory T cells (Tr1) [16]. Regulatory cells and immunosuppressive cytokines present in the intestinal LP and mLN promote tolerance over inflammation in the GALT, on systems homeostasis. Putting all these components together, we have a graph as shown on the figure 1. On the total, we considered 16 humoral and 16 cellular components relevant to the dynamical process, plus the infectious agent: the antigen. From these assembling, we devise a generic model which shall heuristically produce a way to study the oral tolerance considering its complex aspect.

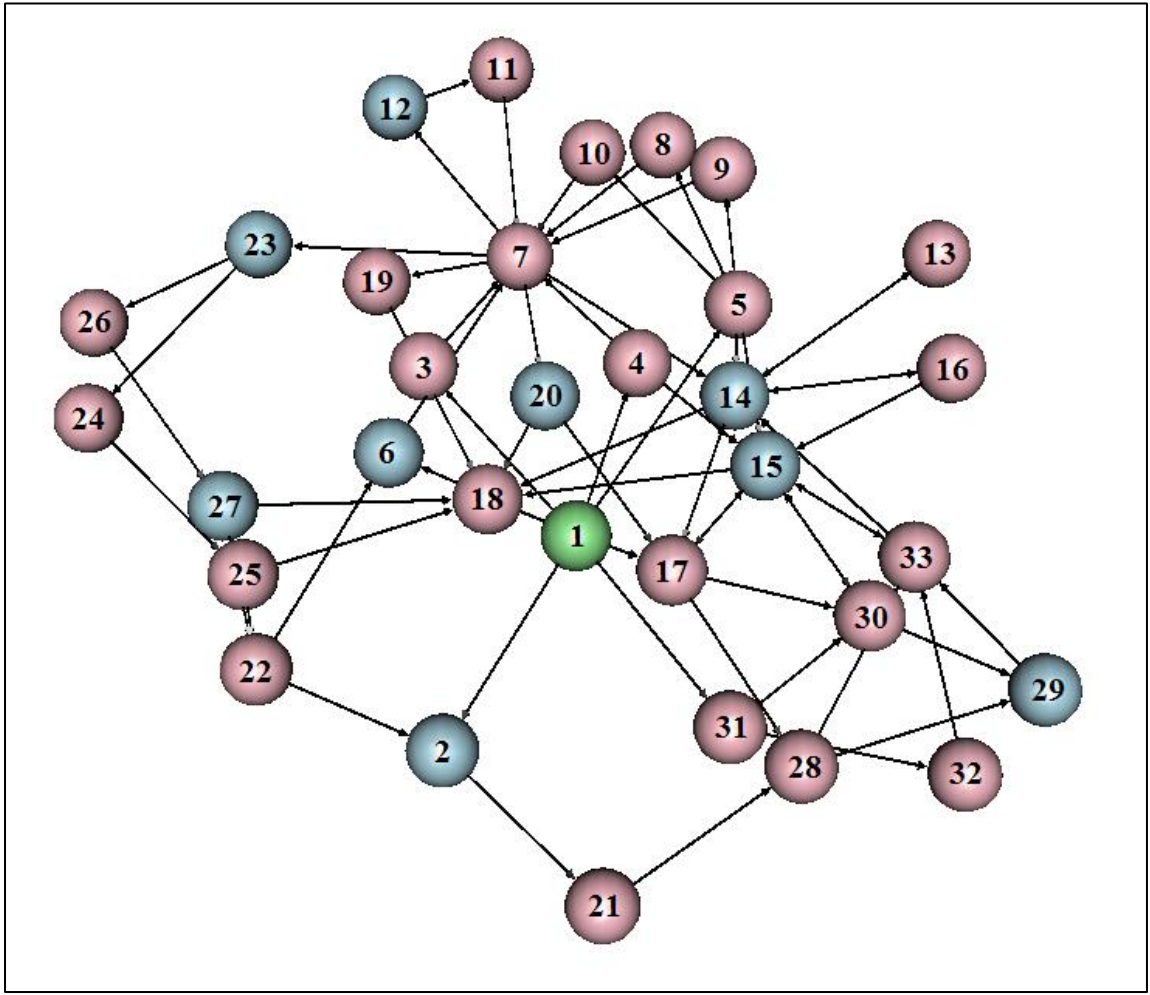

Figure 1. Relational immunological network comprising elements involved on the Oral Tolerance. Pink vertices are related to cellular components and light blue components to humoral components. The following table shows the immunological component related to the numeric label displayed.

| | | |
|---|---|---|
| 1. Antigen | 12. INF-Lambda | 23. RA |
| 2. sIgG | 13. Th3 | 24. α4β7 integrin |
| 3. Goblet Cells | 14. TGF-B | 25. MadCAM-1 |
| 4. CX3CR1 Macrophages | 15. IL-10 | 26. CCR9 |
| 5. Enterocyte | 16. natural TregFoxP3$^+$ | 27. CCL25 |
| 6. sIgA | 17. iTregFoxP3$^+$ | 28. CD11c$^+$ |
| 7. CD103$^+$ | 18. Naïve CD4FoxP3$^-$ | 29. IL-27 |
| 8. Tight junction | 19. CCR7 | 30. Tr1 |
| 9. Transcellular route | 20. IL-2 | 31. M Cell |
| 10. MHC II | 21. FcRn | 32. pCDs |
| 11. CD3$^+$ | 22. B-Cell | 33. CD11b$^+$ |

Table 1. List of the immunological components of the graph displayed as numeric labels on the figured 1 and figure 2.

**The dynamical model**

In this particular work, we propose a new method to understand how the interactive immunological components of the oral tolerance are related to each other. To approach this endeavor, we create a *random walk's algorithm applicable on fixed directed graph's topology*. Overall, this information processing aims to generate a statistical quantity that gives us a notion on how the oral tolerance phenomenon occurs via complex behavior between its immunological components (i. e. cells and humors involved into the immunological process).

Let's consider the immunological interactive network as fixed network, in other words, a constant number of vertices and edges. Assuming that the network is complete, meaning that there is any vertex or portion of the network unconnected to the rest. The connectivity between vertices can be represented by binary values of the adjacency matrix; if there is an edge connecting the vertex *i* and *j*, the value of the adjacency matrix's element $a_{ij}$ will be 1; on the contrary, the inverse is also truth.

As we use directed graph, the following equation is not necessarily achieved

$$a_{ij} = a_{ji} \tag{1}$$

as an artifice of this phenomenon, the degree that we shall treat on this work is the *out degree*, $k_{out}$. Considering a generic vertex *I,* its *out degree* can be written as

$$k_{i\ in} = \sum_i a_{ij} \tag{2}$$

which is actually the sum of the elements of the row *i*. Pictorially, this quantity is expressed into the graph as the number of edges pointing out from the vertex *i*.

As these two topological concepts are established, we can define the *random walk on network* for purposes of our study. In general, the random walk on a complex network can be understood as a stochastic process in discrete time-steps, which a walker follows a path determined by the network's topology, considering his actual position into it. To formalize this process, some rules of "walking onto the network" com de summarized:

I. On a initial time *t*, a generic walker *w*, which is found on the a vertex *i* on the network, chances his position to a vertex *j* as the times chances from *t* to *t+1*;

II. The walkers are allowed to shift positions, from a vertex *i* to any other vertex *j* only if *j* is a first degree neighbor of *i* (i. e. separated only by an edge); given the time chances of *t* to *t+1*;

III. Additionally, the walker must respect the directionality of the edges, given that the network should be directed; if the walk is found on the vertex *i,* he will only go to another vertex *j* if there is a edge point from the vertex *i* to the vertex *j*;

Basically these three rules permeate the random walk which we desire to model our phenomenon. When we introduce more walker and each walker has its position and dynamics, it is of convenience to create a *positional walking vector* $W(g;t)$. Given an initial time *t*, this vector of size $N(t)$ (this function stands for the total walker on the network in the time *t*), and its components are associated to a walker on the network:

$$W(g;t) = \left(w_1(t), w_2(t), w_3(t), \ldots w_{N(t)}(t)\right) \qquad (3)$$

where *g* stands for the graph which the walker are found and the generic position $w_{N(t)}(t)$ is to topological position of the n-esimal walker inserted onto the network. The values of components are the actual position on which the walkers are found to the time being. Concerning a stochastic feature, we may write the probability of position change of a walker as times passes:

$$P_{ij}(t+1) = \frac{a_{ij}}{k_{i\,out}} \qquad (4)$$

We point out that for directed graph, there following asymmetry towards walking probabilities:

$$P_{ij}(t) \neq P_{ji}(t) \qquad (5)$$

This simply means that the probability to a walker to arrives on the vertex *j* from the vertex *i* is usually different from arriving to *i* from *j*. This feature is specially truth as the network is directed, and the *out degree* of *i* and *j* is frequently different. Taking care with this limitation, now we begin to build a way to generate an appreciable statistical quantity that shall grants us immunological insights about the phenomenon. Put this, we introduce the concept of *positional topological vector* which associates each component of it to vertex of the network, and is values are the number of walkers in a given time:

$$S(g;t) = \left(\sigma_i(t), \sigma_j(t), \sigma_k(t) \dots \sigma_n(t)\right) \tag{6}$$

where $\sigma_i(t)$ stands for the number of walkers in the time *t* on the vertex *i*. We can perceive that for each time-step, there will be a contribution of walkers associated to the topological position (vertex). So, if we calculate the *relative value* of each component's value of the above-stated vector *S*, we can write a quantity associated to each vertex that we call *flux of walkers*:

$$f_i(t) = \frac{\sigma_i(t)}{N(t)} \tag{7}$$

as expressed before, $N(t)$ is the total number of walker on the time *t*. To each component of the vector *S* we may associate to a *flux of walker*, and as a new vector we build up:

$$F(g;t) = \left(f_i(t), f_j(t), f_k(t), \dots f_n(t)\right) \tag{8}$$

This vector is of statistical interest to this work, for it is a dynamical quantity dependant on the random choice of each walker on the vertices of the network. Extrapolating this principle, we can set the time as $t \to \infty$, the *flux vector* $F(g;t)$ tends to a stationary distribution; by stationary we mean that for a sufficient large time, the values of the *F* vector won't change. Using this dynamical feature we devise a method to measure the *fluxes* and how the topology affects it.

For instance, if we have a initial graph *g* and we can calculate all the components of the vector $F(g; t \to \infty)$; and, afterwards, we make a topological change on the graph, generating *g'*. To the new graph, we apply the same method to collect the *fluxes*, so we shall have a $F'(g'; t \to \infty)$. The topological change shall reflect a dynamical change, resulting on:

$$F(g; t \to \infty) \neq F'(g'; t \to \infty) \tag{9}$$

The difference of vectors can be used for statistical purposes, as far as the topological modifications $(g \to g')$ implies a dynamical response on the *flux vectors*. To approach this process, we simply calculate the relative error between each component's pair $f_i(g; t \to \infty)$ and $f_i'(g'; t \to \infty)$. We use the time $t \to \infty$ as the both vectors achieve a stationary distribution. This procedure means:

$$\begin{aligned}
f_i(t \to \infty) - f_i'(t \to \infty) &= \Delta f_i \\
f_j(t \to \infty) - f_j'(t \to \infty) &= \Delta f_j \\
f_k(t \to \infty) - f_k'(t \to \infty) &= \Delta f_k \\
&\vdots \\
f_n(t \to \infty) - f_n'(t \to \infty) &= \Delta f_n
\end{aligned} \tag{10}$$

In general, these differences may be negative or positive, given that *f'* can be higher than *f*, or the inverse. For each of these cases, the relative error must obey the following function:

$$p_i = \begin{cases} \dfrac{\Delta f_i}{f_i(t \to \infty)}, & for\ \Delta f_i > 0 \\ \dfrac{\Delta f_i}{f_i'(t \to \infty)}, & for\ \Delta f_i < 0 \end{cases} \tag{11}$$

This function is made necessary in order to scale each case of the variation $\Delta f_i$. As the values of $p_i$ are calculated, we can write the *relative error* as we sum all relative errors divided by the number of vertices:

$$P(g; g'; t \to \infty) = \frac{\sum_{i=1}^{n} p_i}{n} \tag{12}$$

The quantity represented on (12), $P(g;g';t \to \infty)$, é the *relative deviation*; is a quantity which varies from zero to one, where the values close to zero means that the topological modification $(g \to g')$ caused no significant impact to the dynamical structure of the network. When values are close to one, it indicates that the topological modification cause great damage onto the network. Overall, this explains how the dynamical process works, from this we can collect information to create network, and study them by its topological composition. This model has the strong point of allowing us to describe dynamical processes on a network without relying on information about how much a component on the network are related to each other, but rely only on the topology of these relations.

**Algorithm's development and statistics**

The algorithm that comprises all steps described on the model and has the finality of generating a *mean flux vectors* by some fixed *random walking* parameters. The main parameter is the time available to the system to evolve the *walk*. On this particular paper, to each time-step, there will be created a walker and he walks one time as far as his current topological position allows it; keeping in mind the three rules for the *random walking* described above. Additionally, for each time-step, the *positional walking vector* receives a new component with value 1, which means that the created walker is "spawned" on the position 1 in the network. The 1 value means, for our immunological network, represents the presence of a stimulatory antigen in intestinal lumen. After a sufficiently large time ($t \approx 1000$), by the *positional walking vector*, the algorithm generates a *positional topological vector*. So, every site on the network there will be a value associated to it, which is the number of walkers for that component. So it is made a ratio between the number of walkers of that particular component and the total number of walker into the network, this shall be the *flux of walkers*; this is actually the relative frequency for each vertex of the network. This information allows us to built the *flux vector*, that will generate our *standard values* which encompasses the "healthy" network, without any topological disruption.

The secondary parameter of the algorithm is the number of times $L$ that all this process is repeated. The repetition is necessary as far as the *flux vectors,* for each run of the time elapsing process finished, generate different distributions. For our purposes, as the network is relatively small, we used $L = 100$. That for each repetition we get a *flux vector*, so we can write a *mean flux vector* which is the mean value for each component between all vectors:

$$\bar{F}(g;t;L) = \frac{\sum_{L=1}^{L} F(g;t)}{L} \qquad (13)$$

The sum must obey the rules of vector operations. Finally, the *mean flux vector* is our final statistics of interest. So our statistical quantification depends on the time *T*, the number of walkers for a given time *W(t)* and the number of repetitions of the process *L*. So for an initial *graph g*, we get a $\bar{F}(g; t = 1000; L = 100)$ vector, which is actually a frequency distribution. The statistics consists on generating all KOs for vertices; which means generate 32 types of graphs, which one without a vertex. A KO graph consists on the original graph removed a vertex; normally the knockout graph takes the name of the vertex removed from it. As the original "healthy" graph consists on 33 vertex, or 32 knockable vertices (given that the antigen vertex has no meaning for KO), we studied how this removal influenced the *mean flux vector* for each case.

Given these premises, we can compare the values of the *standard graph* and the other *ko graphs* and their respective *mean flux vector*. For each KO, there will be a *mean flux vector* which shall be compared its results in terms of the *standard graph* in order to generate the *mean relative deviation:*

$$\bar{P}(g;g';t \to 1000) = \frac{\sum_{i=1}^{n} \bar{p}_i}{n} \qquad (14)$$

where *g* is the normal graph and *g'* stands for the KO graph. This value $\bar{P}$ shall tell us how the KO was significant for the processes of oral tolerance.

**Results**

After the implementation of the algorithm, we could calculate the *mean deviation* in terms of each simulated *KO graph* and the *standard graph*. The figure 2 displays the values in a graphic. The values of each *mean deviation* can be found on the table 2. It was observed that the knockout of vertex CD103$^+$ caused the greatest impact on network *standard flux*, resulting in a *mean relative deviation* of 0,65. The knockout of vertexes iTregFoxP3$^+$ and Tr1 lead to a mean relative error of 0,37 and 0,38 respectively, and for TGF-β and RA, 0,29 and 0,23 respectively (table 3). In a brief analysis, the results obtained correspond to biological data, where DCs CD103$^+$ plays critical roles in the generation of Tregs, through mechanisms dependent on TGF-β and RA. Tr1 cells acts by the secretion of IL-10, helping to maintain tolerance to commensal antigens [8, 15, 19-21].

| IMMUNOLOGICAL COMPONENT | MEAN DEVIATION | IMMUNOLOGICAL COMPONENT | MEAN DEVIATION |
|---|---|---|---|
| 2. sIgG | 0,221459236 | 18. Naïve CD4FoxP3$^-$ | 0,182795259 |
| 3. Goblet Cells | 0,151888022 | 19. CCR7 | 0,14414499 |
| 4. CX3CR1 Macrophages | 0,118418048 | 20. IL-2 | 0,141569184 |
| 5. Enterocyte | 0,204440683 | 21. FcRn | 0,193757638 |
| 6. sIgA | 0,184631407 | 22. B-Cell | 0,099443526 |
| 7. CD103$^+$ | 0,653053253 | 23. RA | 0,239157237 |
| 8. Tight junction | 0,168152 | 24. α4β7 integrin | 0,139098012 |
| 9. Transcelular route | 0,109349428 | 25. MadCAM-1 | 0,166909826 |
| 10. MHC II | 0,10102181 | 26. CCR9 | 0,138112501 |
| 11. CD3$^+$ | 0,187654376 | 27. CCL25 | 0,149611309 |
| 12. INF-Lambda | 0,12779032 | 28. CD11c$^+$ | 0,162072123 |
| 13. Th3 | 0,14273724 | 29. IL-27 | 0,130526702 |
| 14. TGF-β | 0,293815658 | 30. Tr1 | 0,387507884 |
| 15. IL-10 | 0,198373566 | 31. M Cell | 0,186782946 |
| 16. natural TregFoxP3$^+$ | 0,115992696 | 32. pCDs | 0,098917731 |
| 17. iTregFoxP3$^+$ | 0,374496339 | 33. CD11b$^+$ | 0,170199203 |

Table 2. Immunological component's list, and its numerical label, along with the values of each *mean deviation* impact in relation to the *standard graph*. The values can be interpreted as the topological impact caused on the network.

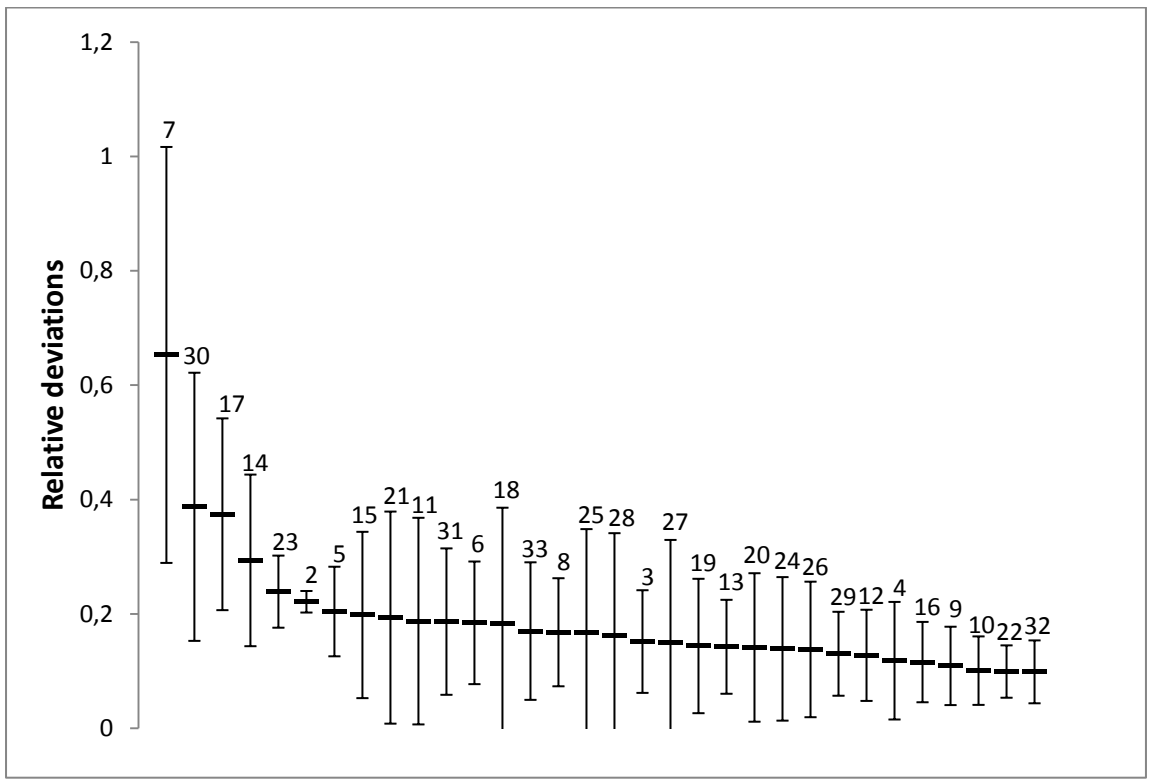

Figure 2. Graphical representation of the mean deviation for each KO. The labels stand for the numeric representation from the immunological elements from the table 1. The errors bars are the standard deviation for each KO.

**Discussion and implications of the model**

Around 40 years ago, Niels Kaj Jerne proposed a new theory to explain the basis of immune system behavior, suggesting the existence of a functional network, centered on patterns of idiotype recognition, carried by lymphocytes [22]. This brief concept was the spark that awakened interest in researchers to not only view the nervous system through models, but also consider the immune system. Notable proprieties of the system as: diversity, self-regulation, memory, connectivity and stability to perturbations, lead to the construction of mathematical models based on differential equations, automaton cellular model and Boolean functions [23-26].

One of the most intriguing proprieties of the immune system is self-regulation; the development of limited responses when constantly perturbed. In this scenario, the gut represents the site where most of antigenic contact occurs in the organism. The adult human gut has nearly 500 to 1000 species of microorganisms, which are in a density of $10^{14}$ cells, corresponding to ten times more the number of somatic human cells and surpassing by two orders the human genetic potential [27]. Beyond the gut microbiota, 130 to 190 g of food proteins are daily absorbed, consisting as another major source of antigenic stimulation in the GALT [28]. However, when in physiological conditions, the prevalent response in the GALT is tolerance. The phenomenon occurs through the multiple action of immune components (dendritic cells, T cells, plamocytes, cytokines) interacting in a unique microenvironment, that is shared by other mucosal sites [6, 7, 10].

In the present work, we investigated the immune components and respective interactions that generate, maintain and regulate the phenomenon of oral tolerance, through the creation of a complex network of interactions. The majority of biological networks are scale-free, displaying the degree distribution of vertex in a power law [29]. In our network, the low clustering coefficient and degree distribution of vertex don't display power law dependence, so there is no characterization as a scale-free network. Although the topology didn't show any relevant feature, we assumed that our network was too little to the expected feature to be displayed. As the phenomenon of our study limited to the oral tolerance, we used a part from the total immunological system, so for topological purposes our network is insufficient.

The results obtained in the *standard flux* suggest that iTregFoxP3$^+$ cells are constantly activated, noting that TGF-β and IL-10 are the two main suppressor cytokines secreted by these cells. Induced Tregs in peripheral tissues play its suppressor role (inhibition of Th1, Th2 and Th17) in a cytokine dependent mechanism, while natural Tregs act preferentially in a cell-cell manner [30]. The high flux of information in *naïve* CD4$^+$FoxP3$^-$ cells indicates a dynamic renewal of T cells in the intestinal *lamina propria* and mesenteric lymph nodes. With the constant entry of antigens (food proteins, microbiota, other environment antigens) in the intestinal luman, new Treg cells are generated (FoxP3$^+$, Tr1, Th3) and even Th17 cells participates in the GALT homeostasis [30-32]. Activated TregFoxP3$^+$ cells induce the secretion of IL-27 by dendritic cells, which stimulate the proliferation of Tr1 cells and production of IL-10. This mechanism positively regulates IL-10 in LP and mLN [16]. The other two main phenotypes of dendritic cells present in the GALT (CD11b+ CD103- e CD11c+ CD11b-) were found in the standard flux. Dendritic cells are essential in the process of oral tolerance, as well as in the immune response to harmful antigens and pathogens. CD11b$^+$ cells produce IL-10, and consistently maintain the Treg population in LP. CD11b$^+$ are also important in the processing of antigens in Peyer's patch [33, 34]. At last, nTregs were activated in the standard flux, maintaining tolerance to self-antigens. Pacholczyk *et al.* (2007) [35] showed that nTregs can cognate non-self antigens, tolerating in this way, commensal antigens. Although they might be dispensable in the induction of oral tolerance [36], the majority of nTregs respond to commensal antigens in the GALT [37]. Briefly, the standard flux found in our model display some remarkable features of oral tolerance; the activation of Treg cells (iTregFoxP3$^+$, Tr1, Th3, nTreg), dendritic cells (DCs) and production of suppressive cytokines TGF-β and IL-10, which occurs by different cell types present in LP (i. e. DCs, Tregs, Macrophages, Enterocyte) [6].

As the network vertices were individually removed (knockout) and *relative deviations* were calculated, we managed to quantify the the relative importance any particular immunological elements involved on the phenomenon. The knockout of vertex CD103$^+$ caused the greatest impact on the network, leading to a relative error of 0,65. DCs CD103$^+$ are the main population of dendritic cells in intestinal LP and mLN. The expression of α4β7 integrin restricts the migration of these cells between LP and mLN, which corresponds to its role in receiving antigens from intestinal lumen and load these antigens to mLN, where they are presented to *naïve*CD4$^+$ T cells [15]. DCs CD103$^+$ produce suppressor cytokines TGF-β, IL-10 and retinoic acid (RA), through the expression of RALDH2, essential for the generation of iTregs. Beyond its synergic action with TGF-β in the generation of iTregs, RA is also important in the synthesis of gut-homing molecules, like the α4β7 integrin and CCR9, allowing Tregs to return from mesenteric lymph nodes to intestinal LP [7]. Mice knockout for CD103$^+$ (CD103$^{-/-}$) showed lower frequency of Tregs and T effectors, significant reduction in the expression of gut-homing molecules in T cells, and oral tolerance induction was impaired [38]. The removal of iTregFoxP3$^+$ and Tr1 resulted in a relative error of 0,37 and 0,38 respectively. FoxP3 is the key regulator of Tregs, being indispensable for its function [30]. Mice null FoxP3$^{-/-}$ developed devastating autoimmune disease, marked by splenomegaly, lymphadenopathy, insulitis, severe skin inflammation, delayed body development and less survival [39, 40]. Yet, this data do not discern between inducible Tregs and natural Tregs, developed in the thymus. However, evidence indicates that iTregs are the main responsible for oral tolerance [30]. The lack of specific markers for Tr1 cells make it hard to comprehend its specific functions and use in clinic. Recently these cells were identified as CD4$^+$CD49b$^+$LAG-3$^+$. Tr1 cells are also induced in the GALT, and respond through the high production of IL-10 [41]. As the knockout of FoxP3, mice that lack LAG-3 exhibit leucocyte infiltrate in multiple organs followed by autoimmune disease [42]. However, the main cytokine secreted by Tr1 cells, IL-10, is dispensable for oral tolerance induction in low doses. Whereas the encephalomyelitis was worse in IL-10$^{-/-}$ mice, the oral administration of myelin glycoprotein of oligodendrocytes (MOG $_{35-55}$) result in improvement of disease in all groups [43]. TGF-β plays distinct functions in the GALT; promotes the expression of FoxP3, regulates Treg function, as well as the polarization of Th17 under the presence of IL-6. As FoxP3 and LAG-3 null mice, mice that lack TGF-β present spontaneous autoimmune disease, and depletion of TGF-βR II of T cells resulted in a similar phenotype, but less aggressive, marked by spontaneous activation of T cells, production of autoantibodies and leukocyte infiltrate in multiple organs [44]. It is worth noting that mice null for TGF-β do not provide a clear view about TGF-β and its function on oral tolerance, since TGF-β takes part in numerous processes linked to organism development, and in its absence, drastic changes occur [45]. In the lack of TGF-β receptors in T cells, less CD103$^+$ cells developed and INF-γ levels were increased [46]. Mice submitted to a vitamin A free diet present dendritic cells positive to langerin in mLN, suggesting that RA creates the unique microenvironment in mLN, different to that found in cutaneous lymph nodes [47].

**Conclusion**

So, we conclude that the model based on complex network is capable in describing the dynamics of immune system in oral tolerance. Our model addresses both topological proprieties and dynamical relations, implemented by a random walk algorithm. The major limitation found in the model is that it lacks a fully quantitative component. Nevertheless, the construction of a qualitative dynamic model for oral tolerance could reflect empirical observations, through the standard flux results and relative error based on individual knockouts. This first model, allied to other models [48, 49] stand as a starting point in the construction of quantitative network models that may describe the kinetics and intensity of casual relationships between the components of the immune system.